# Lithography Free Process for the Fabrication of Periodic Silicon Micro/Nano-Wire Arrays and Its Light-trapping Properties

Divya Rani[1], Anil Kumar[1], Anjali Sain[2,3], Deepika Singh[1], Neeraj Joshi[1], Ravi Kumar Varma[1], Mrinal Dutta[4], Arup Samanta*[1,5]

[1]Department of Physics, Indian Institute of Technology Roorkee, Roorkee-247667, Uttarakhand, India
[2]Photovoltaic Metrology Section, Advanced Materials and Device Metrology Division, CSIR-National Physical Laboratory, New Delhi 110012, India
[3]Academy of Scientific and Innovative Research (AcSIR), New Delhi 110012, India
[4]India National Institute of Solar Energy, Gurgaon 122003, Haryana, India
[5]Centre of Nanotechnology, Indian Institute of Technology Roorkee, Roorkee-247667, Uttarakhand, India
*arup.samanta@ph.iitr.ac.in

**Abstract**
Vertically aligned silicon micro/nanowire arrays of different sizes have been synthesized by combining the modified metal-assisted chemical etching (MACE) and reactive ion etching (RIE) methods. This is a novel lithography-free method to fabricate silicon micro/nanowire arrays. The size of micro/nanowire arrays is controlled by controlling the etching rate and diameter of silica particles. The silicon micro/nanowire geometry can utilize for efficient collection of photo-generated charge carriers from impure silicon wafers, which have a short minority carrier diffusion length also act as a self-antireflection coating layer. For micro/nanowire having average diameters of 40 nm, 330 nm and 950 nm and their corresponding average length of 1.12 µm, 1.1 µm and 1 µm, respectively, the observed average reflectance was 0.22 %, 0.6 % and 0.33 % at 45º incident angle, while the average reflectance was increased up to 4.2 %, 9.2 %, and 11 %, respectively at 75º incident angle in the broad range of 300 - 1200 nm of the solar spectrum. The measured average reflectance for these samples is quite low compared to the planar silicon wafer. Thus this geometry is a promising candidate for fabricating low-cost and highly efficient radial junction silicon micro/nanowire arrays based solar cells.

## Introduction

Silicon is the second most abundant element found in nature after oxygen. With its natural abundance, stability, nontoxic nature, and supportive indirect bandgap, silicon is the most dominant material for commercial electronic applications. Its attractive property as a solar cell absorber layer has ruled the photovoltaics industry until now [1]. On the other hand, the cost of such solar cells is still very high. The major portion of the cost of silicon solar cells is embedded in the silicon wafer. Although metallurgical grade crystalline silicon (c-Si) is inexpensive at only about $1.75-$2.30/kg [2], making efficient solar cells from metallurgical grade c-Si is too far due to the content of many impurities. Due to the presence of impurities, the diffusion length of minority carriers will decrease, which reduces the efficiency of the solar cells. On the other hand, because of the indirect bandgap behavior of c-Si inadequate absorption of light take place at longer wavelengths. This leads to needing for a thick c-Si substrate for high-efficiency solar cells, which also adds a high cost to the solar cells. Thus, device geometry that permits low-cost c-Si and limits the amount of c-Si needed is the vertically aligned micro/nanowire array-based solar cell in radial junction configuration [3]. In the case of micro/nanowire arrays-based solar cells, p-n junction in the radial direction decouples the requirements for carrier collection and light absorption into orthogonal spatial directions [4].

So, each wire length in the array should be optimum in the incident light direction to enhance the light absorption and thin in other dimensions (radial direction) for effective carrier collection [5]. Hence light is absorbed in the axial direction while the collection of charge carriers takes place in the radial direction [4,6]. Also, vertically aligned micro/nanowire array geometry has the advantage of natural incorporation of the antireflection property over a broad spectral range [4,6]. Thus, Si micro/nanowire array geometry possesses unique structural, optical, and electrical properties different from bulk Si. This led to the use of vertically aligned silicon micro/nanowire arrays has also increased in other electronic devices like field-effect transistors [7], high-capacity lithium-ion rechargeable batteries [8,9], chemical [10,11,12], and bio-sensors [13,14,15], particularly in high density integrated electronics [16]. Vertically aligned Si nanowire array also worked as a photocathode for the photo-electrochemical generation of $H_2$ gas [18,19].

There are mainly two approaches, bottom-up and top-down, employed to fabricate size and inter-wire-space controlled vertical Si micro/nanowire array. Vapor–liquid-solid (VLS) growth method is the most popular bottom-up approach to fabricate vertically aligned Si micro/nanowire array. Atwater groups have mainly used the VLS method to fabricate Si micro/nanowire array [20,21]. The main advantage of the VLS growth process is that Si wafers can be easily reused since the wires can be simply transferred to different substrates. However, this method has some undesirable process aspects in the fabrication of micro/nanowire array, which include induced substrate stress due to high temperature (>400°C) process, the method used to deposit metal catalyst of moderate shape, and contamination of silicon with a metal catalyst [22]. Impurities from metal catalysts remain after VLS growth, which is a major reason for low yields and poor performance of semiconductor devices. Dry etching and wet etching are the top-down approaches for fabricating the vertical Si micro/nanowire array. Deep reactive ion etching (DRIE) [23,24], photolithography [24], etc., are commonly used dry etching methods for the fabrication of Si micro/nanowire arrays. In dry etching, Si can be etched independently irrespective of the crystal orientation of the c-Si substrate [24]. However, the above-used methods are expensive, hence limiting their applications. In recent years, metal-assisted chemical etching (MACE) [25,26], a wet etching method, received a lot of interest as it is an inexpensive and high throughput method. By using this method, a high aspect ratio vertically aligned micro/nanowire can be grown with smooth sidewalls [22]. In MACE, metals (e.g., Pt, Ag, and Au) deposited on the c-Si substrate act as a catalyst when submerged in a combination of hydrofluoric acid (HF) and hydrogen peroxide ($H_2O_2$). These metals are deposited on Si substrate by electron beam evaporation, spin coating, electroless deposition, thermal evaporation, or sputtering. After metal deposition, Si oxidized electrochemically, and subsequent HF etching eliminates $SiO_2$. Using the MACE method, Si micro/nanowire arrays of control geometrical parameters like wire diameter, length, and inter wire gap can be fabricated. No requirement of sophisticated instruments makes the MACE method a low cost, and ultimately with this advantage cost of the electronic devices (e.g., solar cells) reduces drastically.

This paper explored the modified MACE method, which is the lithography-free process for developing size and inter-wire-space controlled silicon micro/nano-wire arrays. The following steps have accomplished the fabrication of micro/nano-wire arrays: (i) synthesis of silica nano/micro-particles, (ii) monolayer coating of these particles on silicon substrate by dip-coating method, (iii) control of the inter-particle spacing (50-200 nm) by inductively coupled plasma reactive ion etching method and (iv) metal-assisted chemical etching technique to remove unwanted silicon.

## Experiment

The overall fabrication process is schematically sketched in Figure 1. Initially, silica particles of different sizes 40 nm, 538 nm to 1 micrometer are synthesized using the modified Stöber [27] and Bogush [28] method jointly. In the synthesis process of silica particles, first, the borosilicate beaker and graduated cylinders were cleaned by ethanol, acetone, and Deionized (DI) -water subsequently for 5 min each. A fixed amount of ethanol and ammonium hydroxide ($NH_4OH$) solution is poured into a beaker, and the solution is stirred for 15 minutes at 300 RPM on a magnetic stirrer at room temperature. Then, a fixed amount of DI water is added to the continuously stirred solution. After 15 min, a fixed amount of TEOS was added dropwise into the solution with continuous stirring. The reaction is continued for 8 hours. Finally, we got the only micro/nanoparticle suspension after continuous centrifugation and sonication of colloidal suspension.

After the silica micro/nanoparticles are fabricated, the self-assembled monolayer of the prepared particles is coated on a silicon (Si) substrate (monocrystalline p-type ⟨100⟩) of the size of 1 x 1cm$^2$ by Dip Coating method. Prior to monolayer fabrication, all the samples (Si substrates) are soaked in RCA solution (1 part 27% $NH_4OH$, 1 part 30% Hydrogen peroxide, and 5 parts DI) for 15 minutes to remove the organic residues from the Si wafer. As a result, Si gets oxidized, and a thin layer of thin Oxide is formed on the wafer's surface. After that, samples are cleaned three times with DI water. The native oxide layer is removed from the Si by dipping the substrates in 2 %. HF solution in DI water for 5 minutes and cleaned the samples three times by DI water. After that, we immersed all the substrates in piranha solution ($H_2SO_4$: $H_2O_2$ = 3:1) for 24 hours to make the substrates more hydrophilic and remove any organic surface contaminants. Finally, the silicon substrates are immersed in RCA solution for 30 minutes, rinsed with DI water, and dried under $N_2$ flow. Then Si substrates were coated by NPs suspension (2.5 wt%, sonication time ~6 hours) by Dip coating apparatus with optimized parameters like withdrawal speed ($V_w$)~3mm/sec, the temperature NP suspension ~ 25°C, and immersion time ($t_i$)~ 4 min, and monolayer drying temperature ~ 80°C with a dry time ($t_d$)~2 min. After the monolayer fabrication process, Reactive Ion Etching (RIE) method is used to control the inter-particle spacing between the NPs. In this process, Plasma is created by $CF_4$ gas with a flow rate of 20 sccm RF power of 200W and pressure of 2 Pa at room temperature. In the next step, gold(Au) film is coated using sputtering on the monolayer coated Si substrates as a catalyst. Consequently, nanoparticles were removed from the Si substrates by ultra-sonication in DI for 3-5 minutes. Subsequently, in the final step, micro/nanowire array growth is carried out by etching the samples with an oxidizing etchant, which consists of ($H_2O_2$, 0.44M + DI + HF, 4.6M) in a Teflon container at room temperature for 7minute. After etching, gold (Au) layers are wrapped with sample surfaces. The pretreated samples are treated with Aquaregia (Au etchant solution consisting of $HNO_3$: HCl in the ratio of 1:3 by volume) to remove the Au layer altogether and are rinsed with DI. Morphological analysis of the samples is carried out by Field Emission Scanning Electron microscope (FE-SEM), and reflection measurements are examined by spectroscopic ellipsometry in the 300 - 1200 nm wavelength range.

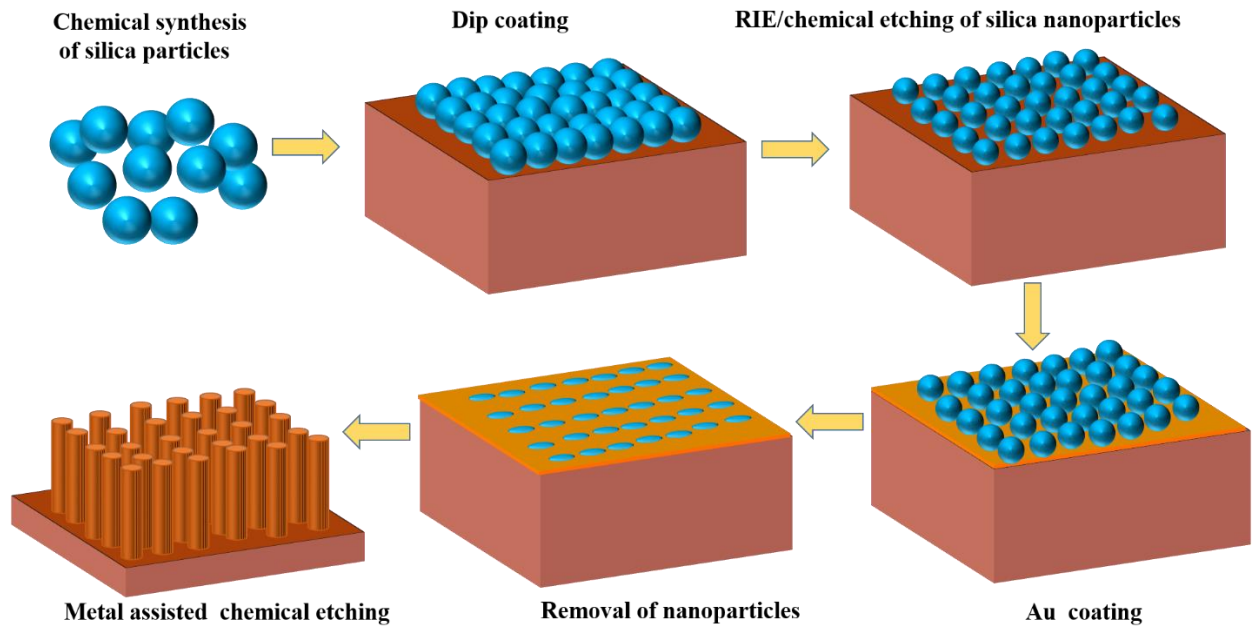

Figure1. Schematic depiction of the fabrication process.

## Results and Discussion:

Figure2 (a), (b), and (c)shows the FE-SEM images and corresponding particle size distribution histogram of synthesized Silica NPs of different sizes ~ 40 nm (0.05M TEOS, 0. 3M NH$_4$OH and 0.1M DI), ~550 nm (0.4M TEOS, 2M NH$_4$OH, and 7M DI), and ~1100 nm (1M TEOS, 2M NH$_4$OH and 7M DI). Silica micro/nanoparticles were synthesized as they work as a masking layer to fabricate the size and inter-wire-space controlled silicon micro/nano-wire arrays. Silica micro/nanoparticles have extra advantages as the synthesis process of their fabrication is very simple and easy to implement without any sophisticated instruments. The preparation of monodisperse silica particles includes the hydrolysis and condensation of alkoxysilanes (often tetraethyl orthosilicate (TEOS)) in a mixture of alcohol, water, and ammonia as a catalyst [17] We observed that on increasing TEOS concentration to an optimized value, silica particles size got to raise.

As the monomer source, TEOS concentration determined the concentration of nuclei/ primary particles present in the system. Aggregation of primary particles results in the formation of secondary particles through Ostwald ripening mechanism. The process continued until a stable condition was achieved, i.e., primary particles consumed. The rise in the primary particles concentration at the induction period results in increased particle size, i.e., [primary particles] ∝ [TEOS] [29]. Thus, particle size has increased as the concentration of TEOS increases.

We have used these synthesized silica micro/nanoparticles for the monolayer fabrication on p-type ⟨100⟩ Si substrates (made hydrophilic) using the dip-coating method. Figure2 (d), (e), and (f) show the FE- SEM images of a self-assembled monolayer of silica micro/nanoparticles of various diameters ~ 40 nm, 550 nm, and 1100 nm. Silica nanoparticles having an average size ~40nm were coated on silicon substrates using NPs suspension concentration ~ 0.5 w/v % and coating parameters $V_w$ ~5 mm/sec, $t_i$~ 20 min, and $t_d$~2 min. While the silica micro/nanoparticles had the size of 550 nm and 1100 nm, they were coated on silicon substrates using NPs suspension concentration ~ 2.5 w/v %, and coating parameters $V_w$ ~ 3mm/sec, $t_d$ ~2 minute, and $t_i$~ 4 min. We observed low defect density, an adequate single layer coverage, and the absence of multilayer silica micro/nanoparticles from these images. The main applications of the coated monolayer of Silica particles are that it acts as an antireflective coating for photovoltaic cells (Solar cells), the lens of glasses, smart coating of vehicles, screens of smartphones, and to create metal nanomesh for Si nanowire synthesis [30]. We have used monolayer coating to create metal nanomesh for Si micro/nanowires synthesis.

After coating the monolayer of Silica NPs on c-Si substrate to control the interparticle spacing between the micro/nanoparticles, Reactive Ion Etching (RIE) method was used, which reduced the diameter of the Silica micro/nanoparticles. Though nanowires with uniform crystallographic orientation can be fabricated by a simple catalytic etching process, the nanowire's size and location cannot be controlled. For device applications, precise control of crystallographic orientation, size, packing manner, and location of nanowires is necessary, which is the most crucial challenge to be overcome.

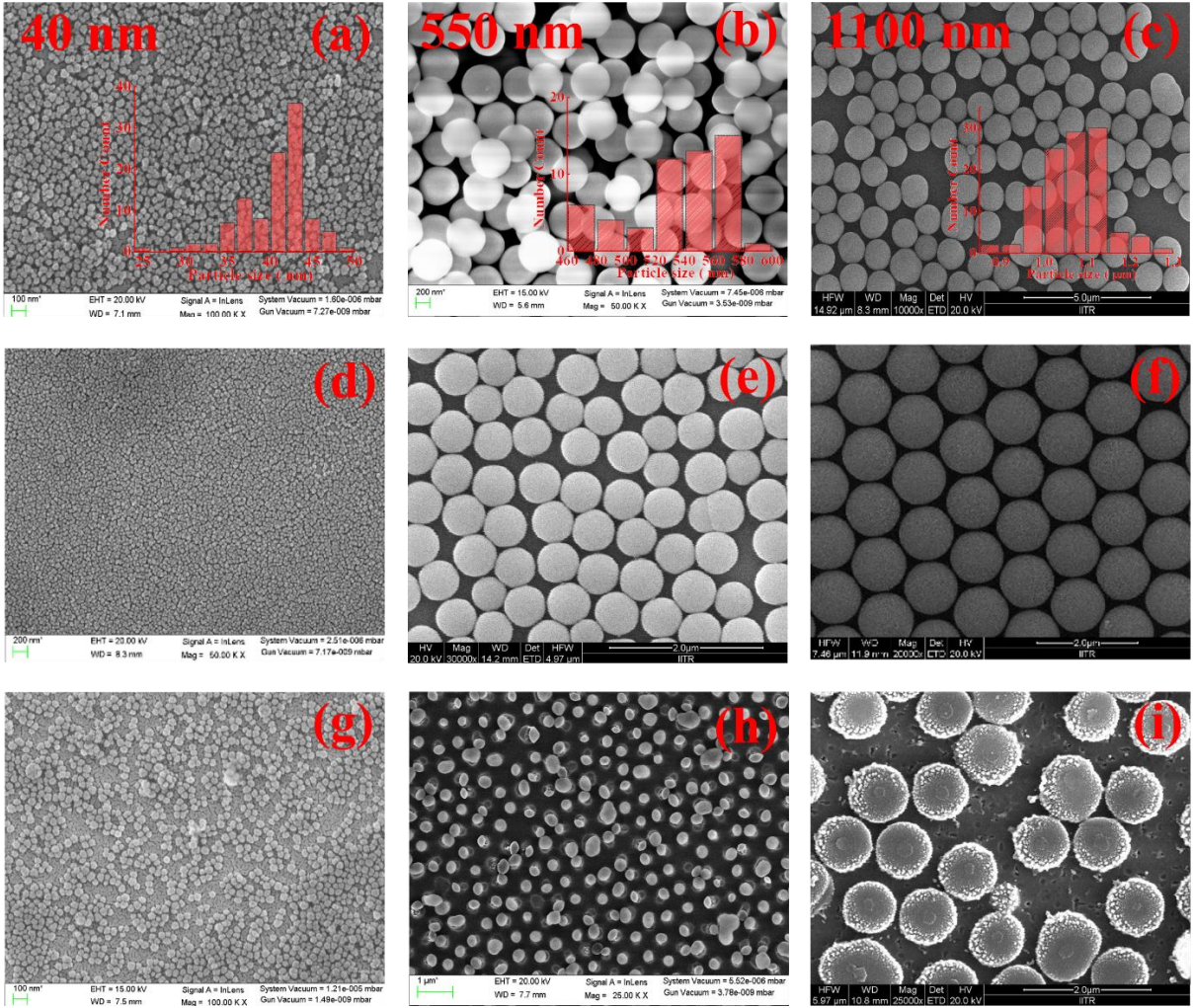

Figure 2: FE-SEM images of the Si substrate surface coated with Silica particles of different diameters comprising (a) 40 nm, (b) 550 nm, and (c) 1100 nm, monolayer of silica NPs of different diameters, containing (d) 40 nm, (using suspension concentration of 0.5 w/v %, and Dip coating parameters - $V_w$ = 5mm/sec, $t_i$ = 20 min and $t_d$ = 2min.), (e) 530 nm, and (f) 1100 nm, (using a suspension concentration of 2.5 w/v %, and Dip coating parameters - $V_w$ = 3mm/sec, $t_i$ = 4 min and $t_d$ = 2min.) coating on Si substrates using Dip-coating method , HF etching Silica nanoparticles have an original average diameter of (g) 40 nm, and RIE etching of Silica nanoparticles have actual average diameters of (h) 700 nm and (i) 1100 nm.

By controlling the size and location of micro/nanoparticles through RIE, the diameter, the length of the micro/nanowire, and the pitch, i.e., center to center distance between micro/ nanowires, can be easily controlled.

Figures 2 (h) and (i) show the FE-SEM images of the Silica micro/nanoparticles, which have average diameters ~ 330 nm and ~ 950 nm with an inter-particle spacing of ~200 nm and ~180nm, respectively, after the RIE treatment for 4 minutes. The RIE treatment altered the closely packed monolayer assembly into a non-closed packed arrangement. The flowing gas in the RIE experiment was $CF_4$ at a flow rate of 20 sccm and a pressure of 2 Pa, and the applied RF power was 20 W. The silica nanoparticles etching rate (uniform etching) of 48-50 nm/min was obtained. For the 40 nm size (Fig. 2 (g)) of nanoparticles to create the inter-particle gap, chemical etching treatment was utilized as RIE was inappropriate for this size of nanoparticles. We have etched the monolayer coated silica nanoparticles with diluted hydrofluoric acid (HF) ~0.008 volume %, and the inter-particle gap of 6.5 nm was obtained.

Next, Au film is coated using the sputtering process on the RIE treated monolayer coated Si substrates as a catalyst. Consequently, micro/nanoparticles were removed from the Si substrates by ultra-sonication in DI for 3-5 minutes.

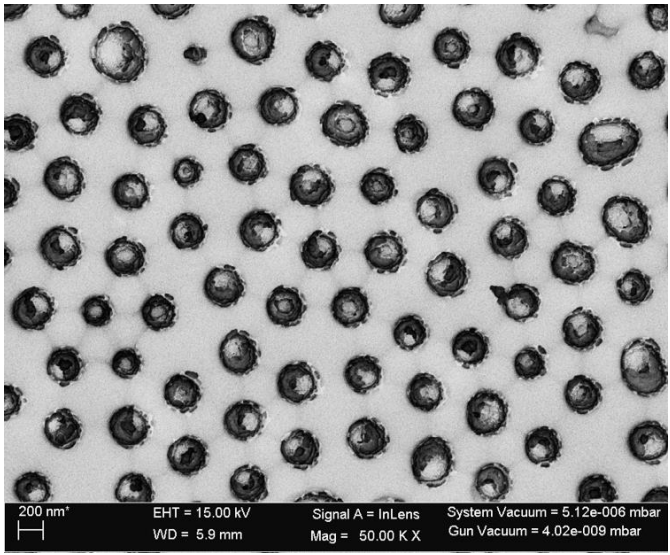

Figure 3. FE –SEM image of the gold-coated sample after removing silica particles ( average diameter ~ 330 nm ).

Figure 3 shows the FE -SEM image of the gold-coated sample after Now there were two regions: Au-coated and the other region from which nanoparticles have been removed, and there was no coating of Au as seen in figure 3. Thus Si wafer was exposed at the position of the silica particles keeping other places gold coated. The region of nanoparticles removed is almost the same diameter as earlier reduced size using RIE. Subsequently, in the final step, the sample was etched by an oxidizing etchant ($H_2O_2$, 0.44M + DI + HF,4.6M) in a Teflon container at room temperature. The etching (etching time ~ 7 min for each sample) took place on the coated area of gold film and left places to stand like wires (the site where micro/nanoparticles have been removed). The diameters of these nanowires were almost similar to the diameters of RIE etched micro/nanospheres. The remaining gold on samples was extracted by aqua regia solution ($HNO_3$: HCl= 1:3). Thus, we obtained the micro/nanowire arrays geometry.

Figures 4 show the FE-SEM images of the Si micro/nanowire arrays fabricated MACE method with RIE treatment. Figures 3(a) and (b), (c) and (d), and, (e) and (f) show the nanowire's front and cross-sectional views fabricated from the sample prepared by Silica NPs having an average diameter of 40 nm, 550 nm and, 1100 nm. We observed that the wires fabricated from the samples prepared by silica micro/nanoparticles of average size ~ 40 nm, ~ 550 nm, and ~ 1100 nm have average diameters of ~ 40 nm, ~330 nm, and ~ 950 nm with their corresponding length ~1.12 µm, ~ 1.1 µm, and 1 µm, respectively. All the figures have variations in length and spacing of the micro/nanowire arrays due to the different sizes of silica particles. There was a gap observed between the nanowires, which is essential for the device fabrication, especially for radial junction solar cells to make the optimum radial junction, which is necessary for the collection of charge carriers. Also, some flat portions were observed in images due to handling while sampling preparation for FESEM and loading time. Such a type of micro/nanowire array-based structure can be beneficial for fabricating solar cells by using lower-grade silicon wafers.

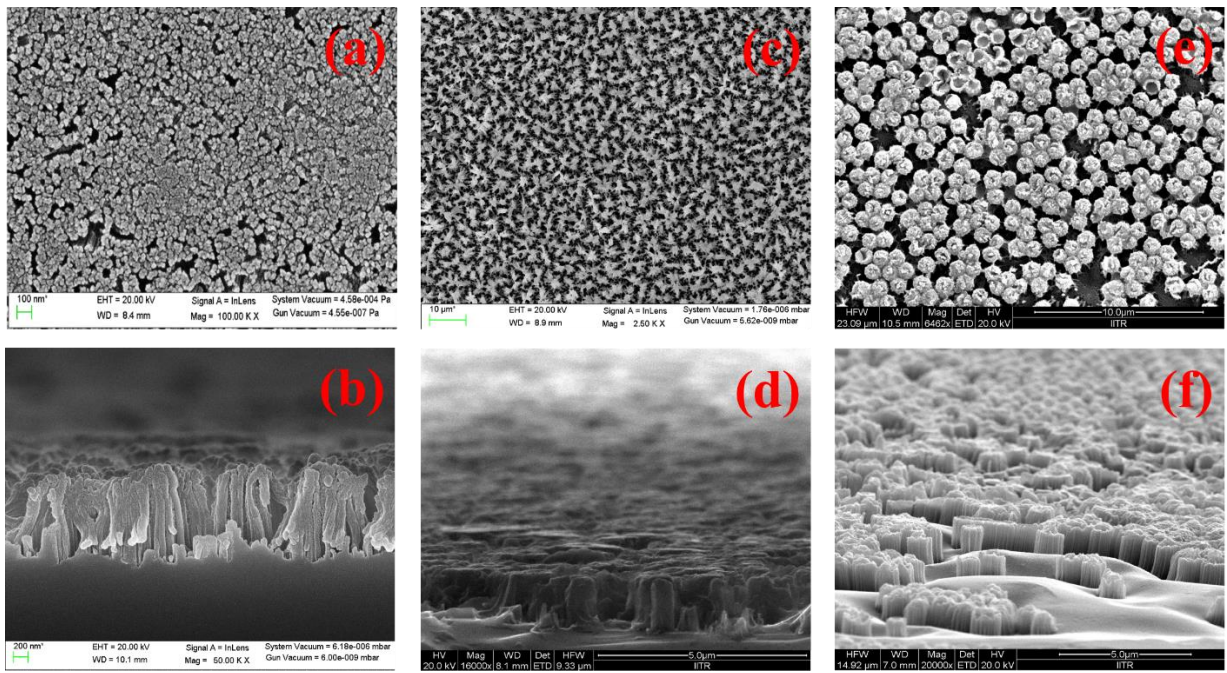

Figure 4. FE-SEM images of the silicon micro/nanowire arrays ((a), (c), and (e) shows the top view, and (b), (d), and (f)shows cross-sectional view fabricating using 40 nm, 550 nm, and 1100 nm silica particles).

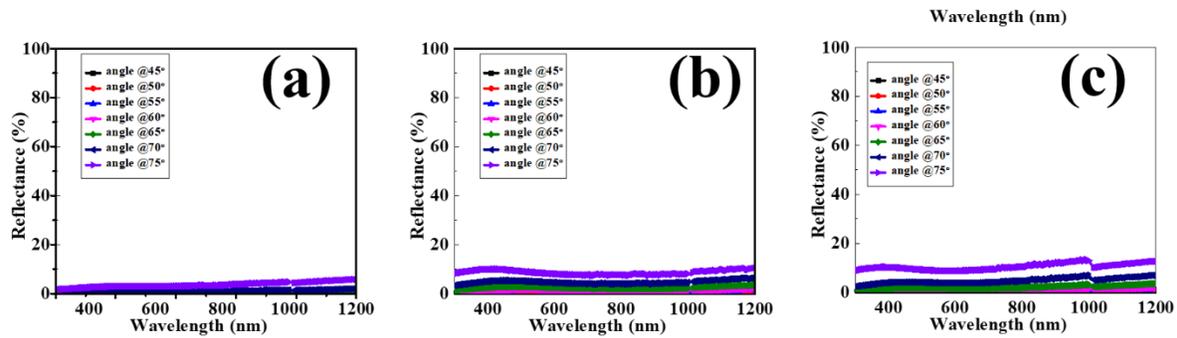

Figure 5. Ellipsometric measurement of the reflectance of the fabricated micro/nanowire arrays for incident angles at 45° to 75° in the interval of 5°: (a) Nano-wires are fabricating using silica particles of average diameter 40 nm, (b) Nano-wires fabricating using silica particles of average diameter 550 nm, and (c) Micro-wires fabricating using silica particles of average diameter 1100 nm.

Figure 5. shows the ellipsometric measurement of the reflectance of the fabricated micro/nanowire arrays. For micro/nanowire having average diameters ~ 40 nm, ~330 nm, and ~ 950 nm and their corresponding average length ~1.12 µm, ~ 1.1 µm, and ~ 1 µm, respectively, the observed average reflectance was ~ 0.22%, ~ 0.6 %, and ~ 0.33%, at 45º incident angle, in contrast, at 75º incident angle, average reflectance was increased up to ~ 4.2%, ~ 9.2 %, and ~ 11%, respectively, in the broad range (300 – 1200 nm) of the solar spectrum. We observed that on increasing the angle of incidence, reflectance got increased. Also, as the size(diameter) of the wires increased, reflectance increased. The maximum value of average reflectance ~11% at 75 ° incident angle was observed for 1µm length of wire arrays which is one-third of the average reflectance (~35%) observed [31] in a planar silicon wafer. For solar cell applications, light reflection from the cell's surface is a significant issue. Thus separate anti-reflection coating is used on the surface of cells, which leads to the high cost of the solar cells; hence this controversy can be solved if we used nanowire/microwire array-based solar cells. Therefore, using this geometry cost of the solar cells can be reduced because by utilizing this geometry, there is no need for a separate antireflection coating on the silicon wafer surface as this geometry has the advantage of trapping light due to multiple absorption and reflection.

## Conclusion

In essence, we have fabricated the silicon micro/nanowire arrays of different sizes by combining the modified MACE and RIE methods. This is a novel lithography-free method to fabricate silicon micro/nanowire arrays. The size of micro/nanowire arrays can be controlled by controlling the etching rate and diameter of silica particles. The minimum value of average reflectance less than 1% was obtained for micro/nanowire arrays in the broad range of the solar spectrum (300 nm-1200 nm) at an incident angle 45°, while the maximum value of average reflectance ~ 11% was observed at incident light 75°, which is one-third of the average reflectance (~ 35 %) of the planar c-silicon wafer. Thus this geometry has the advantage of self-antireflection coating. We can utilize this geometry to fabricate low-cost and highly efficient radial junction silicon micro/nanowire arrays based solar cells.

## Acknowledgment
The authors acknowledge to Ministry of Education and CSIR for fellowship. The work is partially supported by DST-SERB (Project no: ECR/2017/001050), IIT Roorkee (Project no: FIG-100778-PHY), and DST-SERB Ramanujan Fellowship (Project no: SB/S2/RJN-077/2017) India.